\documentclass[aps,prb,twocolumn,showpacs,groupedaddress]{revtex4-1}
\usepackage{amsmath}
\usepackage{eurosym}
\usepackage{graphicx}
\usepackage{dcolumn}
\usepackage{bm}
\usepackage{color}
\usepackage{longtable}
\usepackage{multirow}

\begin{document}

\title{Coexistence of ferromagnetism and superconductivity in CeO$_{0.3}$F$_{0.7}$BiS$_{2}$}

\author{J. Lee$^{1,2}$}
\author{S. Demura$^{3}$}
\author{M. B. Stone$^{2}$}
\author{K. Iida$^{1}$}
\author{G. Ehlers$^{2}$}
\author{C. R. dela Cruz$^{2}$}
\author{M. Matsuda$^{2}$}
\author{K. Deguchi$^{3}$}
\author{Y. Takano$^{3}$}
\author{Y. Mizuguchi$^{3,4}$}
\author{O. Miura$^{4}$}
\author{D. Louca$^{1}$}
\author{S.-H. Lee$^{1}$}

\affiliation{$^1$Department of Physics, University of Virginia,
Charlottesville, VA 22904, USA} 
\affiliation{$^2$Quantum Condensed Matter
Division, Oak Ridge National Laboratory, Oak Ridge, Tennessee 37831-6393,
USA} 
\affiliation{$^3$National Institute for Materials Science, 1-2-1,
Sengen, Tsukuba, 305-0047, Japan} 
\affiliation{$^4$Department of Electrical
and Electronic Engineering, Tokyo Metropolitan University, 1-1,
Minami-osawa, Hachioji, 192-0397, Japan}

\date{\today }

\begin{abstract}
Bulk magnetization, transport and neutron scattering measurements were
performed to investigate the electronic and magnetic properties of a
polycrystalline sample of the newly discovered ferromagnetic superconductor,
CeO$_{0.3}$F$_{0.7}$BiS$_{2}$. Ferromagnetism develops below T$_{FM}$ = 6.54(8) K and superconductivity is found to coexist with the ferromagnetic state below T$_{SC}$ $\sim $ 4.5 K. \ Inelastic neutron scattering measurements reveal a very weakly dispersive magnetic excitation at 1.8 meV that can be explained by an Ising-like spin Hamiltonian. \ Under
application of an external magnetic field, the direction of the magnetic
moment changes from the c-axis to the ab-plane and the 1.8 meV excitation splits into two modes. \ A possible mechanism for the unusual
magnetism and its relation to superconductivity is discussed.
\end{abstract}

\pacs{61.05.F-,74.25.N-,75.25.-j,75.30.Ds,75.30.Kz,75.40.Cx,75.40.Gb,78.70.Nx%
}
\maketitle


\section{Introduction}

The recently discovered BiS$_{2}$-based superconductors\cite{Bi4O4S3,LaOBiS2,NdOBiS2,PrOBiS2,CeOBiS2} 
share many common characteristics with other unconventional superconductors, such as the cuprates and iron-based superconductors. 
Prevalent in theses system is the presence of a square lattice in their layered structure and superconductivity induced by doping charge carriers.\cite{Paglione,Fujita,Bi4O4S3,LaOBiS2} 
\ The superconducting (SC) mechanism in the new BiS$_{2}$-based superconductors is still under debate. 
While electron-phonon coupling constant calculations\cite%
{Phonon1,Phonon2,Phonon3} yield a T$_{SC}$ close to the experimental value
suggesting a conventional phonon mediated mechanism, no significant change
in the phonon density of states has been observed across the
superconducting transition of LaO$_{0.5}$F$_{0.5}$BiS$_{2}$ system\cite{JSLee}. In contrast, other experimental 
\cite{nonBCSE1,nonBCSE2,nonBCSE3,nonBCSE4} and theoretical results \cite%
{nonBCST1,nonBCST2,nonBCST3} suggest an unconventional superconducting
mechanism may exist in this system. \ Conventional superconductivity
with s-wave Cooper pairing would be destroyed in the presence of magnetism
by the orbital effect \cite{Ginzburg} and/or the paramagnetic effect \cite%
{Clogston, Chandrasekhar}. \ Therefore, studying the relation of
superconductivity to magnetism would provide important information on the
nature of the superconducting mechanism. \ 

CeO$_{1-x}$F$_{x}$BiS$_{2}$ exhibits ferromagnetism and superconductivity at
low temperatures \cite{CeOBiS2}, thus providing a good system to investigate
the superconducting mechanism of the BiS$_{2}$ materials. The coexistence of
ferromagnetism and superconductivity deserves attention on its own because
actual systems exhibiting such coexistence are quite rare. \ Examples
include some heavy fermion superconductors \cite%
{HF1,HF2,HF3,HF4,HF5,HF6,HF7,HF8,HF9}, Ruthenate-layered cuprates 
\cite{Ru1,Ru2,Ru3,Ru4,Ru5,Ru6}, Eu(Fe$_{1-y}$Co$_{y}$)$_{2}$(As$_{1-x}$P$%
_{x} $)$_{2}$ \cite{Eu1,Eu2,Eu3,Eu4,Eu5,Eu6}, and CeFe(As$_{1-y}$P$_{y}$)(O$%
_{1-x} $F$_{x}$) \cite{Ce1,Ce2,Ce3,Ce4}.

Layered superconductors are typically composed of an alternate stacking of superconducting layers and blocking layers\cite{Bi4O4S3,LaOBiS2,Kamihara,Armitage}. The superconducting layers serve as a conducting path of charge carriers\cite{Kamihara} which becomes superconducting below the transition temperature, and the blocking layers are insulating spacers sandwiched between superconducting layers. The interplay between the superconducting and blocking layers
and its impact on superconductivity has been of great interest in cuprates
and pnictide superconductors. \ Recently the R$_{e}$T$_{m}$P$_{n}$O system (R%
$_{e}$: rare earth, T$_{m}$: transition metal, P$_{n}$: Pnictogen) \cite%
{CeTMPnO1} has shown diverse electronic and magnetic properties depending on
the T$_{m}$P$_{n}$ blocking layers. Experimental systems in this category includes a ferromagnetic Kondo system CeRuPO\cite{CeTMPnO2}, a correlation-enhanced local moment antiferromagnet CeNiAsO \cite{CeTMPnO3} and superconducting LaFePO \cite{CeTMPnO4}. Recent research examining LnOBiS$_{2}$ systems (Ln=La,Nd,Ce,Pr,Yb)\cite{LaOBiS2,NdOBiS2,PrOBiS2,CeOBiS2,LnOBiS2} reveal that the BiS$_{2}$ layer is the common essential component responsible for inducing superconductivity. Density functional calculations\cite{Bi4O4S3,nonBCST1} also find Bi-$6\mathit{p}$ and S-$3\mathit{p}$ bands close to the Fermi level supporting the idea of BiS$_{2}$ superconducting layers. Therefore, it is believed that, in CeO$_{1-x}$F$_{x}$BiS$_{2}$, the blocking layers are formed by magnetic Ce ions separating the superconducting BiS$_{2}$ layers.\ It is important to examine how the two layers interact with each other.

In this paper, the crystal and magnetic structures and magnetic fluctuations
in the superconducting CeO$_{1-x}$F$_{x}$BiS$_{2}$ are examined as a
function of temperature and magnetic field. \ Even though the crystal
symmetry is \textit{P4/mmm}, broadening of the nuclear Bragg peaks is
present indicating poor crystallinity. \ Below T$_{FM}$ $\approx $ 6.5 K,
the Ce$^{3+}$ magnetic moments align ferromagnetically along the c-axis, and
a spin-wave mode around $\hbar \omega $ $\approx $ 1.8 meV appears. The
spin-wave can be described by a 3D Ising spin Hamiltonian with
nearest-neighbor and next-nearest-neighbor interactions.\
Application of a magnetic field leads to a spin flop to the ab-plane while
the $\hbar \omega $ $\approx $ 1.8 meV mode splits into two excitations. These atypical
changes with applied magnetic field could be explained by considering the Ising nature of the Ce magnetic moment together with imperfect crystallinity. We suggest that CeO$_{1-x}$%
F$_{x}$BiS$_{2}$ is a ferromagnetic superconductor where rare-earth layers
show Ising ferromagnetism with negligible interaction with superconducting
layers enabling the coexistence of two typically antagonistic phenomena.

\section{Experiment}

A 1.0 g polycrystalline sample of CeO$_{0.3}$F$_{0.7}$BiS$_{2}$ was
prepared with a solid-state reaction and then annealed under high-pressure using a Cubic-Anvil-type high-pressure machine in air condition. Bi$_{2}$S$_{3}$ powders were obtained by sintering the mixtures of Bi grains and S grains in the evacuated quartz tube at 500 °C for 10 hours. Mixtures of Bi grains, Bi$_{2}$S$_{3}$ grains, Bi$_{2}$O$_{3}$ powders, BiF$_{3}$ powders and Ce$_{2}$S$_{3}$ powders with nominal compositions of CeO$_{0.3}$F$_{0.7}$BiS$_{2}$ were ground, pelletized, and sealed into an evacuated quartz tube. The tube was heated at 800 °C for 10 hours. The obtained pellets were ground and annealed at 600 °C for 1 hour under a hydrostatic pressure of 3 GPa.\
The neutron scattering measurements were performed at the High Flux Isotope Reactor using the HB2A powder diffractometer\cite
{HB2A}, and at the Spallation Neutron Source using the Cold Neutron Chopper
Spectrometer (CNCS)\cite{CNCS}. The samples were loaded into vanadium cans
for the diffraction measurements at HB2A and into an aluminum can for the
inelastic measurement at CNCS under a He atmosphere and mounted on a
cryostat. \ The neutron diffraction data from HB2A were collected at a
constant wavelength of 1.5408 $\text{\AA }$ at 2 K and 20 K. The inelastic
neutron scattering (INS) measurements at CNCS were performed with an
incident energy E$_{\text{i}}\ =$ 4 meV. \ For inelastic measurements, the background was determined and subtracted from the data using an empty can measurement.

\section{Susceptibility and Resistivity}

\begin{figure}[htbp]
\centering
\includegraphics[width=1.00\hsize]{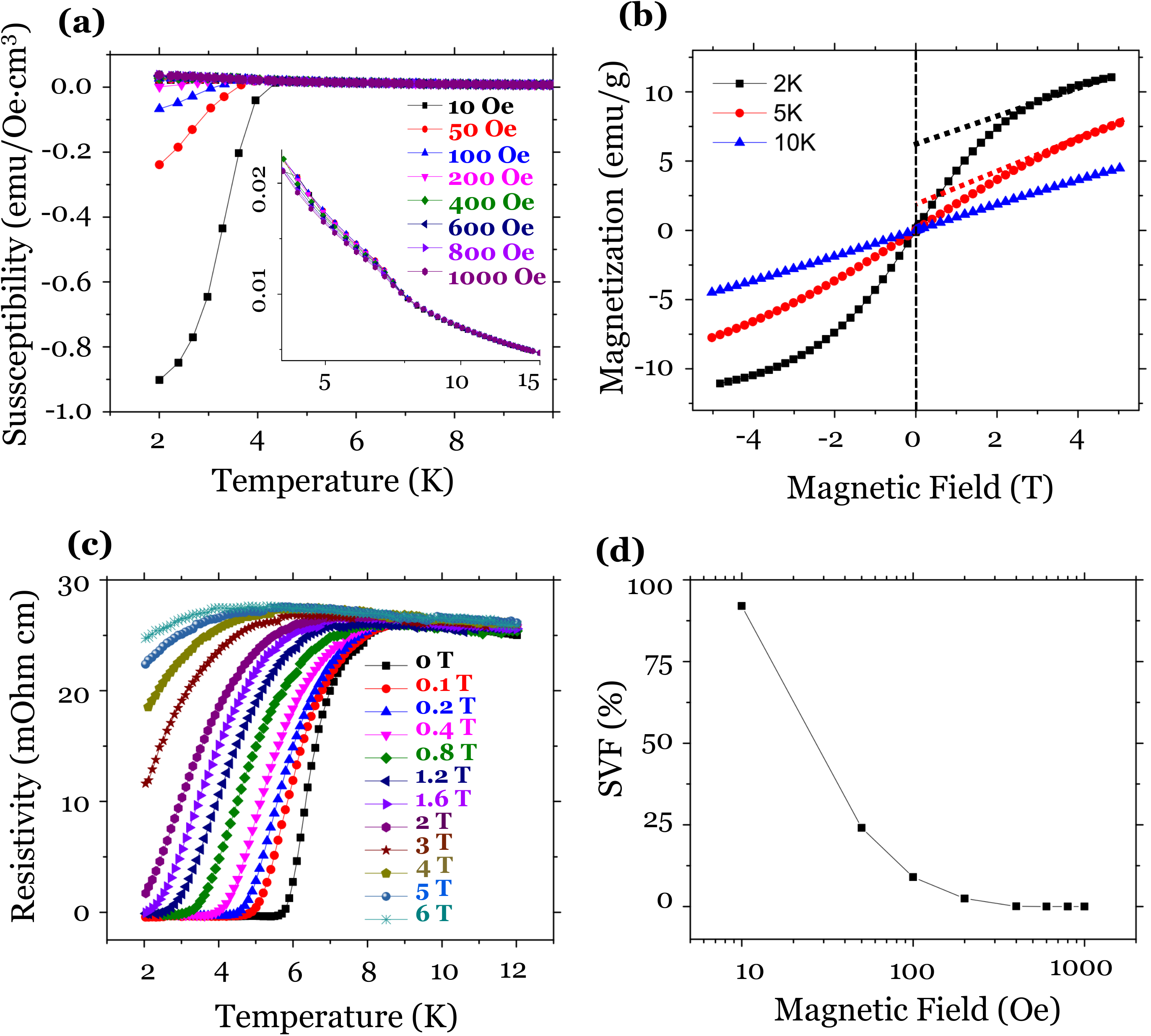}
\caption{(color online) Field dependence measurements of magnetic susceptibility and
resistivity of CeO$_{0.3}$F$_{0.7}$BiS$_2$. (a) magnetic susceptibility with ZFC and FC at various fields. The inset shows an enlarged plot of FC data showing the onset of ferromagnetism. (b) isothermal M-H curve at different temperatures. Dashed lines are linear fits for magnetic fields greater than 3 T. The inset shows enlarged graph showing hysteresis curve coming suggesting ferromagnetic nature. (c) resistivity measured at various fields. (d) estimated SVF as a function of applied magnetic field. }
\label{fig:Fig_BULK}
\end{figure}

Shown in Fig. \ref{fig:Fig_BULK} are the low-temperature bulk magnetization
and transport property measurements as a function of an external magnetic field.\ Figure \ref{fig:Fig_BULK} (a) shows DC magnetic susceptibility with zero field cooling (ZFC) and field cooling (FC) at various fields ranging from 10 Oe to 1000 Oe.\ Upon cooling the magnetic susceptibility, $\chi $, gradually increases at T $\approx$ 8 K, indicating the development of ferromagnetism consistent with the previous report\cite{CoexPRB}. The linear extrapolation of the high magnetic field data to 0 Tesla is shown in Fig. \ref{fig:Fig_BULK} (b), which gives an estimate of the frozen moment of 0.52 $\mu_{B}$/Ce and 0.11$\mu _{B}$/Ce at 2 and 5 K, respectively.\ The magnetic moment is quite smaller than what is expected from $4\mathit{f}^{~1}$ electron of Ce$^{3+}$. \cite{Blundell}

Further cooling leads to a rapid drop in $\chi $ at T$_{sc}$ $\approx $ 4.5 K due to the diamagnetism from the Meissner effect where the system exhibits superconductivity (Fig. \ref{fig:Fig_BULK} (a)). When the external magnetic field is increased, superconductivity is suppressed as evidenced by the weakening of the diamagnetic effect. Fig. \ref{fig:Fig_BULK} (c) represents resistivity measured at various field ranging from 0 T to 6 T. The suppression of superconductivity with increasing field is also observed from non-zero resistivity above 1.6 T. \ Fig. \ref{fig:Fig_BULK} (d) shows the
superconducting volume fraction (SVF) as a function of field, estimated by
the following equation: SVF $\left[ \%\right] =4\pi \times \left\{ \chi
\left( T_{sc}^{onset}\right) -\chi \left( 2K\right) \right\} \times 100$
where $T_{sc}^{onset}$ is the temperature below which the magnetic
susceptibility starts to drop. At zero field, the SVF is over 90 $\%$, which is consistent with a recent specific heat measurement \cite{LnOBiS2}. Increasing field rapidly decreases the SVF, and the superconductivity almost vanishes around 400 Oe, suggesting the zero resistivity between 400 Oe and 1.6 T shown in Fig. \ref{fig:Fig_BULK} (c) is from filamentary superconductivity.

\section{Static spin correlations}

\subsection{Diffraction results}

\begin{figure}[tbhp]
\centering
\includegraphics[width=1.0\hsize]{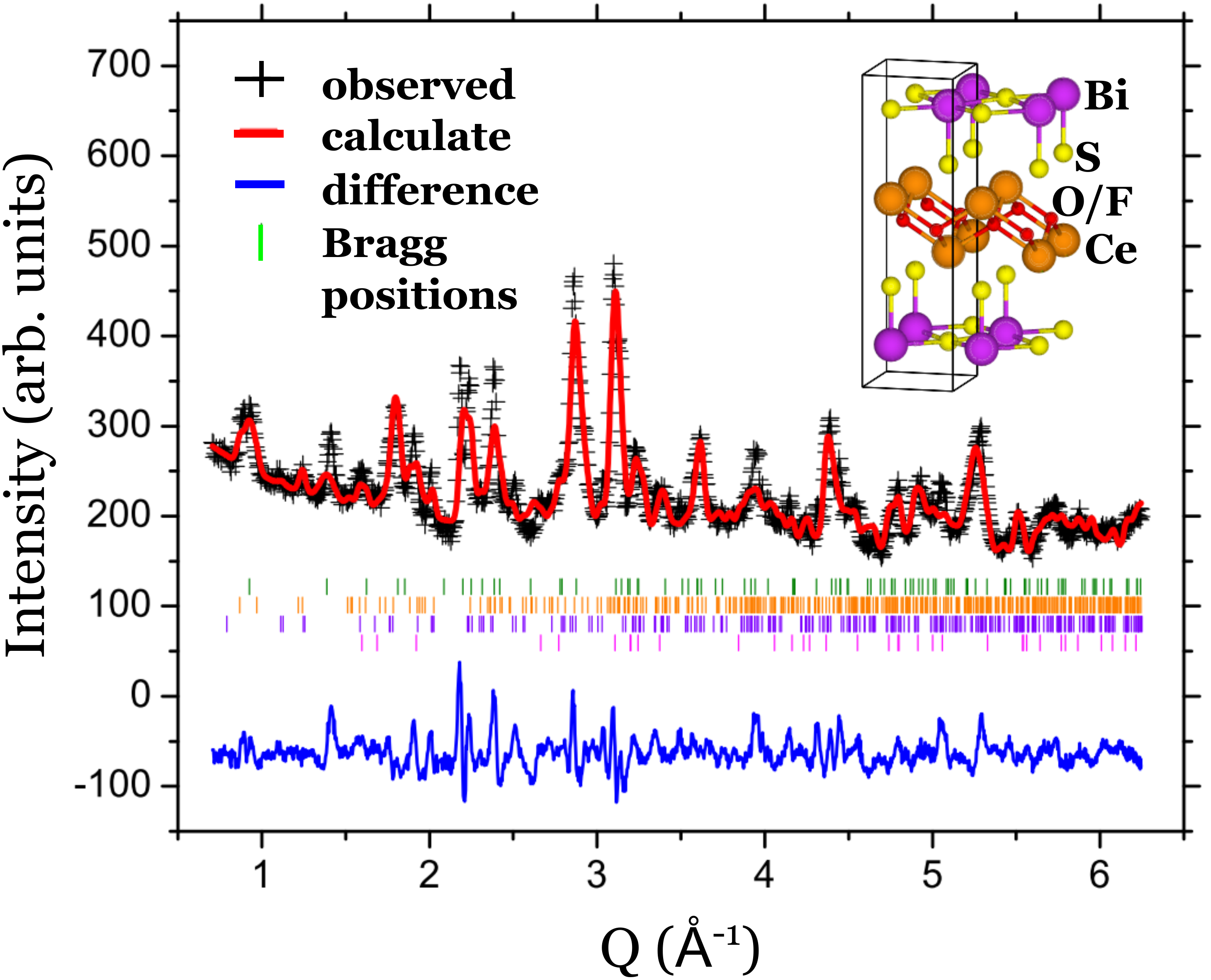}
\caption{(color online) Neutron powder diffraction data from CeO$_{0.3}$F$_{0.7}$BiS$_2$ at T = 10 K. The black crosses are the measured scattering intensity, and the red solid line represents the Rietveld refinement fit to
the data. The vertical bars indicate Bragg reflection positions of the main phase and impurity phases: CeO$_{0.3}$F$_{0.7}$BiS$_2$, CeF, Bi$_{2}$S$_{3}$, and Bi in descending order. Their weight fractions are 99.3(31)$\%$, 0.44(2)$\%$, 0.21(1)$\%$, and 0.06(1)$\%$, respectively. The blue solid line shows the difference between measured and fitted intensities.}
\label{fig:Fig_NPD_NUC}
\end{figure}

\begin{table}[htbp]
\centering
\caption{Refined structural parameters of CeO$_{0.3}$F$_{0.7}$BiS$_{2}$
obtained from neutron powder diffraction using Fullprof \protect\cite%
{Fullprof}. Numbers in parentheses correspond to one standard deviation in
the mean value.\\}
\label{tab:cryst_structure}
\begin{tabular}{c|c|c}
\hline\hline
\multicolumn{2}{c|}{CeO$_{0.3}$F$_{0.7}$BiS$_2$} &  \\ 
\multicolumn{2}{c|}{P4/nmm T=20K} & $\chi^{2}$ = 4.72 \\ \hline
\multicolumn{2}{c|}{a ($\text{\AA }$)} & 4.039(1) \\ 
\multicolumn{2}{c|}{c ($\text{\AA }$)} & 13.566(7) \\ \hline
~~atom~~ & ~~Wyckoff position~~ & z \\ \hline
Ce & 2c (0.25, 0.25, z) & 0.104(1) \\ 
Bi & 2c (0.25, 0.25, z) & 0.614(2) \\ 
S$_1$ & 2c (0.25, 0.25, z) & 0.360(1) \\ 
S$_2$ & 2c (0.25, 0.25, z) & 0.848(4) \\ 
O/F & 2a (0.75, 0.25, 0) & $-$ \\ \hline\hline
\end{tabular}%
\end{table}

\begin{table}[htbp]
\caption{The anisotropic phenomenological strain parameter used to fit the
diffraction pattern of CeO$_{0.3}$F$_{0.7}$BiS$_2$. The numbers in the
parentheses represent estimated errors.\\}
\label{tab:aniso_broaden}
\centering
\begin{tabular}{ccccc}
\hline
S$_{400}$ & S$_{004}$ & S$_{220}$ & S$_{202}$ &  \\ \hline\hline
8.465E+01 & 9.913E+00 & 1.561E+02 & 1.144E+01 &  \\ 
(0.750E+01) & (0.978E+00) & (0.414E+02) & (0.649E+01) &  \\ \hline
\end{tabular}%
\end{table}

The neutron powder diffraction pattern shown in Fig. \ref{fig:Fig_NPD_NUC}
is from data collected on HB2A at T = 10 K. \ Significant Bragg peak
broadening is observed indicating that the system is not very crystalline. \
A similar observation was previously made in a related system, La(O,F)BiS$%
_{2}$ \cite{JSLee}, however the peak broadening is larger in CeO$_{0.3}$F$_{0.7}$BiS$_2$.\ At the same time, several impurity phases are present in the sample, with
some unknown. These unknown impurities are unlikely to compromise our result as their peaks are temperature-independent, thus not belonging to the magnetic phase. Also considering that their summed integrated intensity corresponds to the 1.4(6) percent of total integrated intensity, it is likely that the volume fraction of those foreign phases are less than a few percent.
\ The structural parameters obtained from the refinement are
summarized in Table \ref{tab:cryst_structure}. \ The intrinsic peak broadening does not allow for a full refinement including thermal parameters. \ The main features were indexed using the \textit{P4/nmm}
space group, and the peaks were fit using an anisotropic broadening function 
\cite{Broadening}. \ A set of possible non-zero anisotropic strain
parameters are S$_{400}$=S$_{040}$, S$_{202}$=S$_{002}$, S$_{004}$, and S$%
_{220}$, whose refined parameters are summarized in Table \ref%
{tab:aniso_broaden}.\ The red solid line shown in Fig. \ref{fig:Fig_NPD_NUC} represents the best fit. The inset in the figure shows the crystal structure obtained.\ Superconductivity occurs in the BiS$_{2}$ layers\cite{Bi4O4S3,nonBCST1} where the Bi ions form a square lattice and separated by the Ce(O/F) blocking layer.
\begin{figure}[tbhp]
\centering
\includegraphics[width=1.0\hsize]{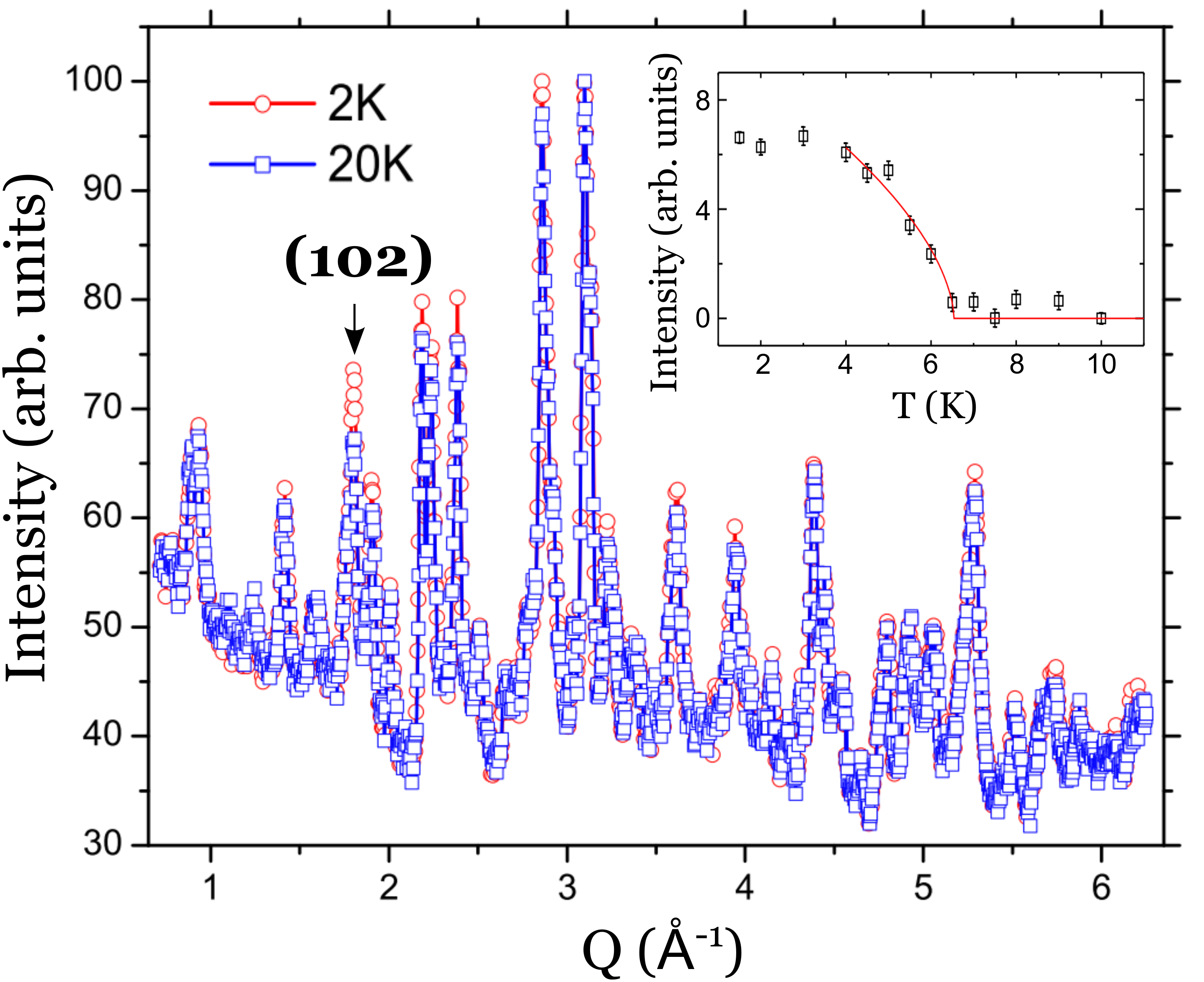}
\caption{(color online) The neutron powder diffraction data of CeO$_{0.3}$F$_{0.7}$BiS$_2$ below and above the magnetic
phase transition temperature, T$_M$. Red circles represent 2K data and blue
squares represents 20K data. Inset shows the summed intensity of the Q =
(102) Bragg peak as a function of temperature. Red line represent the power
law fit down to 4K.}
\label{fig:Fig_NPD_MAG}
\end{figure}

As the sample is cooled to $\approx $ 1.5 K, an enhancement of the
neutron scattering intensity superimposed on the nuclear Bragg intensity is
observed (Fig. \ref{fig:Fig_NPD_MAG}). \ Given that the intensity appears at
the nuclear peaks, it is most likely ferromagnetic in nature with a magnetic propagation
vector of k = (000). The increase in intensity is especially significant at
the (102) peak, and its integrated intensity is shown in the inset as a
function of temperature. \ Upon cooling, CeO$_{0.3}$F$_{0.7}$BiS$_{2}$
undergoes a second order phase transition around T$_{FM}$ $\approx $ 7 K. \
We fit the integrated intensity with a power law above 4 K: I $\propto $ $\left(
T_{FM}-T\right) ^{2\beta }$. \ A T$_{FM}$=6.54(8) and $\beta $ = 0.30(7) are
obtained from this comparison. The value of the critical exponent, $\beta $, is
closest to the theoretical value of a three-dimensional Ising model\cite%
{Collins}, which is 0.326, but given the large error bar, a
three-dimensional XY model ($\beta $=0.345) is also possible.

Given that the nuclear Bragg peaks are very broad and the magnetic signal is week, a magnetic structure refinement is not straight forward. \ This is
especially difficult in a ferromagnet since the magnetic intensity can be
obscured by thermal and/or strain/stress broadening. \ Group theoretical
analysis is used to determine the possible symmetry consistent type of magnetic order in this system. In
the \textit{P4/nmm} crystal space group, there are in total four possible
irreducible representations (IRs) compatible with k=(000): $\Gamma _{2}$, $%
\Gamma _{3}$, $\Gamma _{9}$, $\Gamma _{10}$. They represent
antiferromagnetic ordering with spins along the c-axis, ferromagnetic
ordering with spins along the c-axis, ferromagnetic ordering with spins in
the ab-plane, and antiferromagnetic ordering with spins in the ab-plane,
respectively.

\begin{figure}[tbhp]
\centering
\includegraphics[width=1.0\hsize]{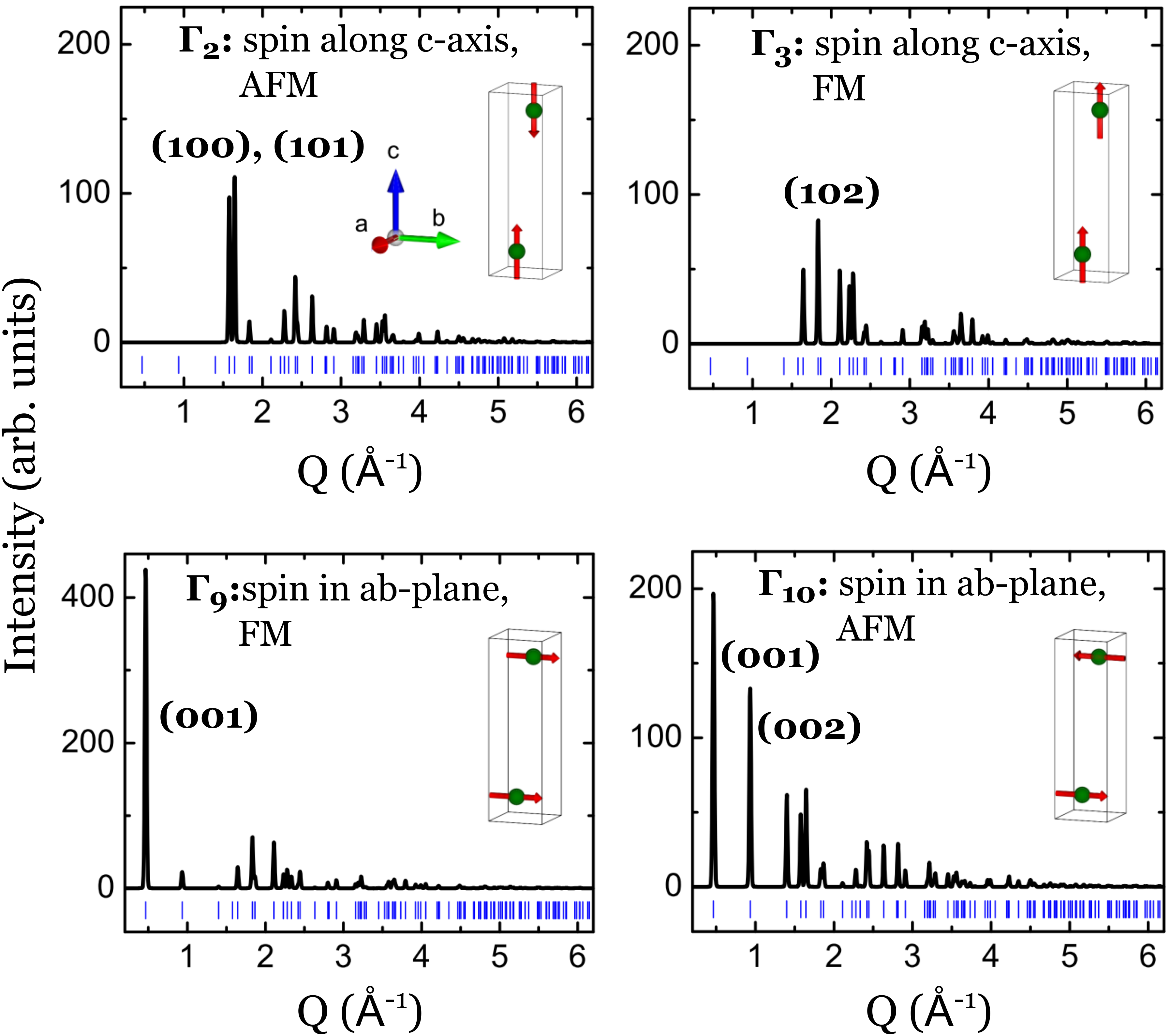}
\caption{(color online) Group theoretical analysis on the possible magnetic structures and
their simulated neutron scattering intensities. Most prominent Bragg peaks
are indicated in each panel. Group analysis was done using the program SARAh%
\protect\cite{SARAh} and the number of IRs follows Kovalev's notation. Insets illustrate the corresponding magnetic order of the Ce moments for each IR.}
\label{fig:Fig_NPD_SIMUL}
\end{figure}

The model magnetic neutron patterns for each IR together with their
corresponding spin configurations are shown in Fig. \ref{fig:Fig_NPD_SIMUL}.
In the case where spins are lying along the c-axis such as in $\Gamma _{2}$
and $\Gamma _{3}$, there cannot be any (00L) magnetic Bragg peaks because
only spin moments that are orthogonal to the wave vector can contribute to
the scattering intensity. \ The strongest peaks are the (100) and (101) in
the antiferromagnetic spin configuration of $\Gamma _{2}$, while the (102)
peak is the strongest peak in the ferromagnetic spin configuration of $%
\Gamma _{3}$. On the other hand, if the spins are in the ab-plane as in $%
\Gamma _{9}$ and $\Gamma _{10}$, (00L) peaks are allowed. In the
ferromagnetic configuration of $\Gamma _{9}$, the (001) magnetic Bragg peak
is clearly the strongest peak. In the antiferromagnetic configuration
of $\Gamma _{10}$, the (001) is still the strongest peak while the (002)
peak is comparable in intensity.%

Since the increase of the measured magnetic scattering is prominent at (102), it can
be deduced that the magnetic structure is of the $\Gamma _{3}$ type, as it
shows ferromagnetic ordering of the Ce$^{3+}$ magnetic ions along the
c-axis. It is unlikely for other IRs to be mixed with $\Gamma _{3}$ as a
small component along the c-axis, whether AFM or FM, would have produced
significant scattering at (001), which is not observed in the data.

\subsection{Elastic measurements}

\begin{figure}[tbhp]
\centering
\includegraphics[width=1.0\hsize]{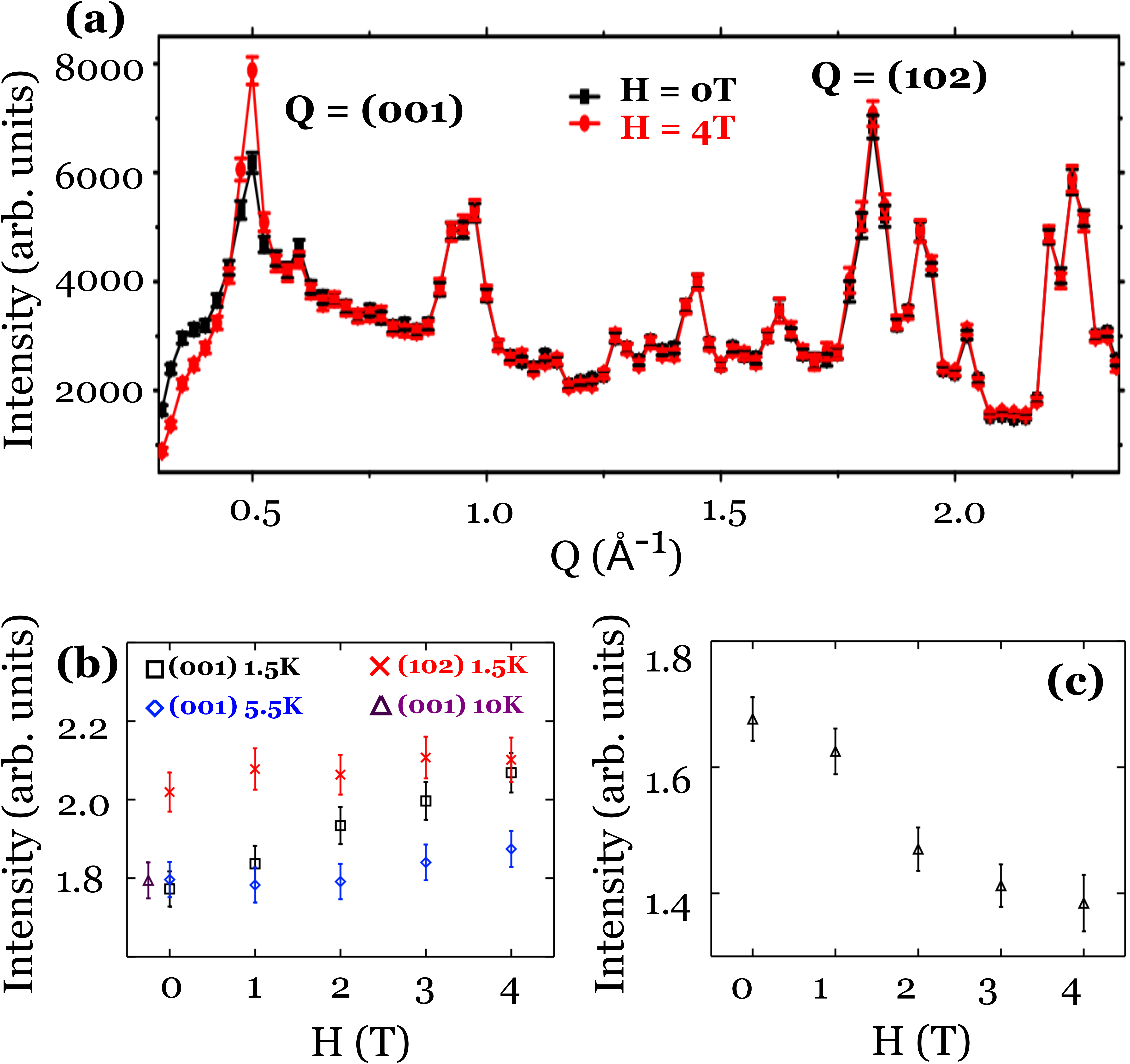}
\caption{(color online) (a) Elastic neutron scattering intensities at different fields at T = 1.5 K from the CNCS measurements. The data were integrated between -0.5 meV and 0.5 meV energy transfer. (b) The field dependence of the summed intensity of the Q = (001) Bragg peak at T = 1.5 K and T = 5.5 K and Q = (102) peak at T = 1.5 K.\ The Q = (001) intensity at T = 10 K at zero field is shown to demonstrated that Q = (001) intensity is purely nuclear at zero field. This intensity is displaced slightly along the negative x-axis direction for visualization. (c) The field dependence of the quasi-elastic scattering intensity at low Q around [0.3,0.46] $\AA^{-1}$ summed over E = [0.1,0.4]meV measured at T = 1.5 K. }
\label{fig:Fig_HDEP_NPD}
\end{figure}

Measurements at CNCS allowed us to reach a lower Q value which showed the
presence of the (001) nuclear peak at 0 T as shown in Fig.\ref{fig:Fig_HDEP_NPD} (a). \ No difference was observed at the two temperatures measured for zero applied field for T = 1.5 and 10 K, which suggests that the intensity under this peak is nuclear.

As shown in Fig. \ref{fig:Fig_BULK} (a), an external magnetic field suppresses
superconductivity. Figure \ref%
{fig:Fig_HDEP_NPD} (a) shows the $Q-$dependence of the elastic neutron
scattering intensity integrated over $\hbar \omega $ from -0.5 to 0.5 meV at
T = 1.5 K with varying field. It can be seen that the ferromagnetic intensity at the (102) Bragg peak does not show significant change with field. \ On the other hand, a notable effect under field is observed with the appearance of an elastic magnetic signal at the $(001)$ reflection. In addition, a broad intensity peak is observed at low Q, below 0.5 \AA $^{-1}$, that shows little temperature dependence and becomes less pronounced with field as in Fig. \ref{fig:Fig_HDEP_NPD} (c). \ This is likely related to the reduced crystallinity of the sample that may give rise to domains with some domains having short-range correlations. \ The integrated intensity of the (001) Bragg peak as a function of magnetic field is shown in Fig. \ref{fig:Fig_HDEP_NPD} (b) at T = 1.5 K $<T_{FM}$ and at 5.5 K just below T$_{FM}$. \ A similar behavior under field is observed at both temperatures, where the intensity increases gradually up to 4 T. The field-induced $(001)$ magnetic peak indicates that the spin configuration changes from the ferromagnetic alignment along the c-axis, $\Gamma _{3}$, at zero field to ferromagnetic alignment in the ab-plane, $\Gamma _{9}$ under applied magnetic field.

\begin{table}[htbp]
\caption{The magnetic Bragg peak intensity ratio of Q = (001) and Q = (102) at different applied magnetic fields. For each field, the expected ratio of ab-component magnetic moment and c-component moment are calculated. For nuclear backgrounds, Bragg peak intensities at T = 1.5 K and T = 10 K were used for Q = (001) and Q = (102), respectively, both at zero field. The numbers in the
parentheses represent estimated errors.\\}
\label{tab:Iratio}
\centering
\begin{tabular}{ccc}
\hline
H (T) & I$_{(001)}$/I$_{(102)}$ & m$_{ab}$/m$_{c}$ \\ \hline\hline
0 & 0 & 0 \\ 
1 & 0.36(39) & 0.25 \\ 
2 & 0.99(59) & 0.44 \\ 
3 & 1.1(5) & 0.47 \\ 
4 & 1.5(6) & 0.57 \\ \hline
\end{tabular}%
\end{table}

We can estimate the ratio of in-plane magnetic moment and c-axis magnetic moment by comparing the intensity of Q = (001) and Q = (102) as in Table III. A gradual increase of in-plane magnetic moment with field is observed. It should be noted that when the moment direction rotates from the c-axis to the ab-plane and the size of moment is fixed, the (102) intensity should decrease. On the other hand, if an ab-plane moment is induced while the c-axis moment remains fixed, the (102) peak will increase. Our measurements as shown in Fig. \ref{fig:Fig_HDEP_NPD} (c) indicate slight increase but only within error bar. The details regarding how much the intensity will change are determined by the amount of induced ab-plane moment and decrease of c-axis moment. It is, however, hard to extract the absolute size of magnetic moment from our neutron diffraction data due to large broadening. 

\section{Inelastic measurements}

\subsection{Spin fluctuation}

\begin{figure}[tbhp]
\centering
\includegraphics[width=1.0\hsize]{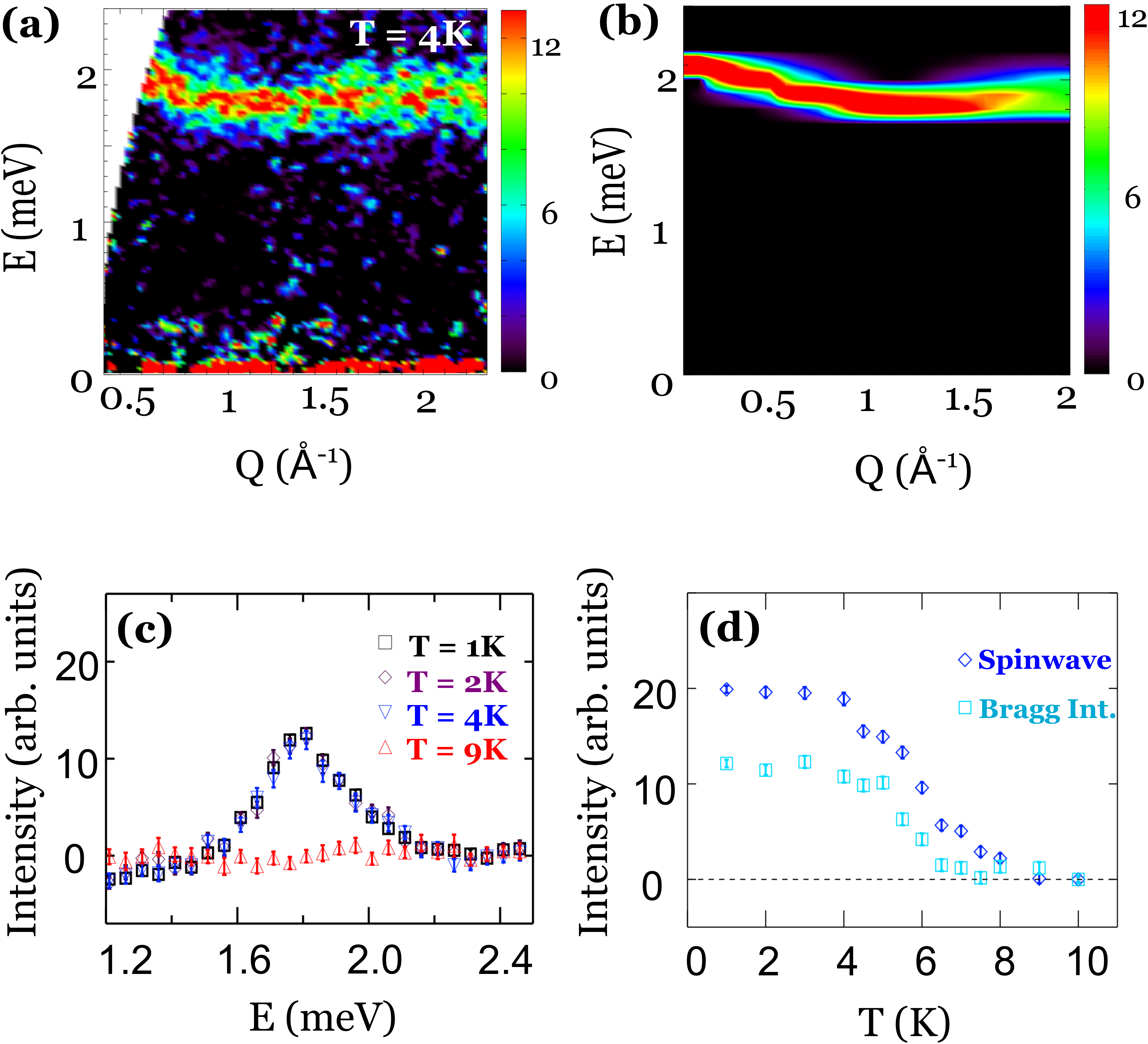}
\caption{(color online) (a) The contour map of measured inelastic neutron scattering intensity 
in reciprocal wave vector (Q) and energy transfer (E) space. 
(b) Simulated spin excitation spectrum in Q-E space.
(c) Inelastic neutron scattering intensities along energy transfer at
different temperatures. Intensities shown are summed over 0.5  $\text{\AA }^{-1}$ $\leq$ Q $\leq$ 1 $\text{\AA }^{-1}$. 
(d) Temperature dependence of integrated intensities around
magnetic excitation and magnetic Bragg peak Q = (102). The $T-$dependence of
the magnetic excitation was obtained by integrating the intensity over $Q$
from 0.5 to 2.0 $\text{\AA }^{-1}$ and over energy transfer from 1.5
to 2.5 meV. All figures shown here are background subtracted by the data obtained at T = 10K.}
\label{fig:Fig_INS}
\end{figure}

In order to investigate how the magnetic correlations evolve through the
magnetic phase transition at $T_{FM} \sim 6.5$ K, inelastic time-of-flight neutron
scattering measurements were performed at several different temperatures
spanning $T_{FM}$. To see the magnetic signals below $T_{FM}$ more clearly,
the T = 10 K $> T_{FM}$ data was subtracted as a background in our analysis.
At 1.4 K, we find the emergence of a strong and flat excitation
centered at $\hbar\omega = 1.8$ meV, as shown in Figure \ref{fig:Fig_INS}
(a) and (c). Upon warming, the 1.8 meV excitation stays flat up to about 4 K
and starts decreasing in intensity to vanish above 8 K (see Figure \ref{fig:Fig_INS}
(d)). The $T-$dependence is similar to that of the FM Bragg peak intensity,
indicating the 1.8 meV mode is most likely a ferromagnetic spin wave.

The simplest Hamiltonian that reproduces all important characteristics of
the spin excitation is the anisotropic exchange spin hamiltonian;

\begin{equation}
H = \sum_{i,j} J_{i,j} \left\{ \alpha {S_{i}^{x}} {S_{j}^{x}} +
\alpha {S_{i}^{y}} {S_{j}^{y}} + {S_{i}^{z}} {S_{j}^{z}}
\right\} + g \mu_{B} \sum_{i} \vec{B} \cdot \vec{S_{i}}
\label{eq:Hamiltonian0}
\end{equation}
where $J_{i,j}$ and $B$ are the exchange integral and the external magnetic
field, respectively. $\vec{S_{i}}$ is the spin operator at the position $r_{i}$,
 and  ${S_{i}^{\nu}}$ represents the ${\nu}$ = x, y, z component of the spin. 
$g$ is the Land$\acute{e}$ g-factor, and $\mu_{B}$ is
the Bohr magneton. The first term describes anisotropic interactions between
the magnetic moments where the degree of anisotropy is controlled by $\alpha$%
, and the second term gives rise to the Zeeman effect from the external
magnetic field. $\alpha<1$, $\alpha>1$, and $\alpha=1$ cases correspond to
Ising-like, XY-like, and Heisenberg spins, respectively. For the exchange
interactions, nearest neighbor(NN) and next-nearest neighbor(NNN) have been
considered. The bond length for the inter-layer $J_1$ and that of the
intra-layer $J_2$ coupling are 4.010(30) $\text{\AA }$ and 4.038(1) $\text{%
\AA }$, respectively.

This Hamiltonian can be solved analytically, and the eigen-energies are the
following, 
\begin{equation}
\begin{split}
\epsilon =& - 4 J_{1} S - 4 J_{2} S + 2 \alpha J_{2} S \left( cos k_{x} + cos k_{y}
\right) \\
& \pm 2 \left| \alpha J_{1} S \right| \sqrt{1 + cos k_{x} + cos k_{y} + cos
k_{x} cos k_{y}} \\
& + g \mu_{B} B_{z} .
\end{split}%
\end{equation}
\
Here, $k_{x}$, $k_{y}$, and $k_{z}$ correspond to the Miller indices of a reciprocal wave vector.\
The theoretical neutron scattering intensities were calculated and averaged
over all solid angles to simulate the data obtained from the polycrystalline
sample. The best result was obtained with parameters, $J_1$ = $%
J_2$ = -0.24 meV with $\alpha$ = -0.1, as shown in Figure \ref%
{fig:Fig_INS} (b). These values suggest that the main interaction between
magnetic moments is ferromagnetic and very Ising-like. We note that the Ising nature is consistent with the result we obtained from
power-law fitting of the magnetic order parameter. The negative value of
anisotropy parameter, i.e., antiferromagnetic x-y interactions, was needed
to produce the up-turn of the excitation mode at low $Q$.

\subsection{Field dependence }
\begin{figure}[htbp]
\centering
\includegraphics[width=1.0\hsize]{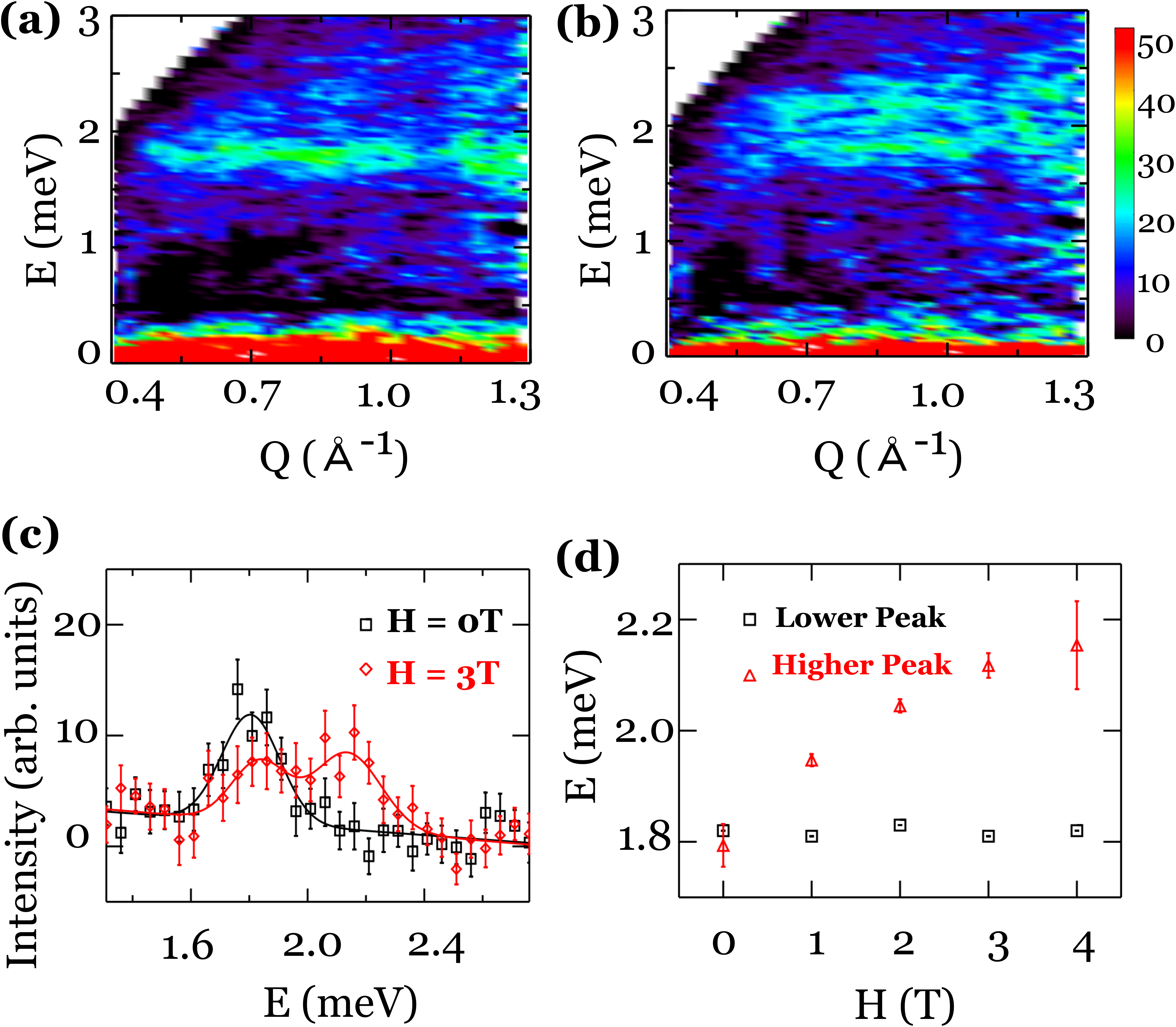}
\caption{(color online) The change of inelastic neutron scattering with external magnetic
field is shown. (a) Spin excitation spectrum at H = 0 T. (b) Spin excitation
spectrum at H = 3 T. (c) Inelastic neutron scattering intensities along energy transfer at different field of 0 T and 3 T. Scattering intensities
are integrated over 0.5 $\text{\AA }^{-1}$ $\leq$ Q $\leq$ 1 $\text{\AA }^{-1}$. (d) Gaussian fitted peak
positions at different magnetic fields.}
\label{fig:FIG_HDEP_INS}
\end{figure}


The magnetic excitations change as a function of applied field as shown in Fig. \ref%
{fig:FIG_HDEP_INS} (a) - (c). The H = 3 T data (Fig. 7 (b)) shows that the
magnetic field smears out the 1.8 meV mode. The energy cut of the data from 1.3meV to 2.8meV shows that under field, the 1.8 meV mode splits into
two peaks, one of which remains nearly at the same starting energy, while
the other peak moves to higher energy as the external field increases (Fig.
7c). This behavior is clearly seen in Figure \ref{fig:FIG_HDEP_INS} (d) where we plot the Gaussian peak positions (see Fig. 7 c) as a function of H at T = 1.5 K.

\section{Discussion}

In the absence of an external magnetic field, the spins in CeO$_{0.3}$F$_{0.7}$BiS$_{2}$  are aligned along the c-direction with a strong
Ising-like exchange anisotropy. The application of field introduces a magnetic component in the ab-plane suggesting that there exists a strong ab-plane magnetic susceptibility compared to a weak c-axis susceptibility. A similar anisotropic magnetic
susceptibility was previously observed in a Ce-based
intermetallic compound, CeAgSb$_{2}$ \cite{CeAgSb2_1,CeAgSb2_2}. The
latter system has a tetragonal crystal structure as well and shows
ferromagnetic ordering along the c-axis below T$_{c}$ = 9 K. This system shows a linear increase of the in-plane magnetization when H // (100), whereas the magnetization along the c-axis remains the same when H // (001). After
intense debate, it has been determined that the magnetic state of
this system can be described by spins in the $\left\vert \pm
1/2\right\rangle $ crystal electric field (CEF) split ground state with anisotropic ferromagnetic exchange interactions \cite{Debate1,Debate2,Debate3,Debate4}. 
The larger in-plane component of $J_{x,y}$ than $J_{z}
$ of the ground state is responsible for the anisotropic susceptibility.
The magnetic moment of the ground state arises from $g_{J}\mu _{B}J_{z}$ =
0.41 $\mu _{B}$, and this value is close to our magnetic moment estimated
from Fig. \ref{fig:Fig_BULK} (b). Considering the same crystallographic space group, ferromagnetic structure,
small magnetic moment, and Ising nature of spin interaction, we believe the
same physics can apply to our system as well. In this picture, the increase
of in-plane magnetic moment comes from the grains where the external field
is applied in the ab-direction. Future experiment with a single crystal
sample would be necessary to confirm this scenario.

The split of the ferromagnetic spin wave under external magnetic field
is quite unusual for a monoatomic ferromagnet that has only one spin wave
mode. Under an external magnetic field, a single spin wave dispersion is
expected to shift by the Zeeman energy, $g \mu_{B} H S$, if the moments
are aligned parallel to the field. Thus, the origin of the $H$-induced
splitting of the 1.8 meV mode into two is likely due to the presence of several magnetic
domains.

A possible explanation for the origin of the split can be a spin-glass-like disorder of the transverse spin components arising from local
atomic distortions with spontaneous magnetization along the c-axis. Such a magnetic
ground is called asperomagnetism \cite{Armophous0} and is usually found in
amorphous crystal structures such as Fe$_{100-x}$B$_{x}$ \cite{Armophous1} or
CeNi$_{0.4}$Cu$_{0.6}$ \cite{Armophous2}. Strong random local anisotropy with a wide distribution of the exchange interactions can be a sufficient condition for the onset
of asperomagnetism. It should be noted that the large broadening of the
nuclear Bragg peaks shown in Fig. \ref{fig:Fig_NPD_NUC} indicates a strong
lattice disorder. The broad background near Q = (001) in which the
intensity is reduced under the magnetic field is also consistent with the spread of
magnetic moments around the c-axis. The fluctuation energy of the
perpendicular spins will not change with the external magnetic field and these moments are likely the ones contributing to the lower-field-insensitive mode.

Lastly, the coexistence of ferromagnetism and superconductivity in this system deserves special attention. From our measurements, we did not observe any change in either the magnetic structure or excitation upon entering superconducting phase. The spin
Hamiltonian most applicable to our system suggests that the inter-layer coupling is negligible. This leads to the suggestions that the coexistence of ferromagnetism and superconductivity is made possible by the electronic separation of the superconducting layer and the magnetic layer, for instance, as in the case of EuFe$_{2}$As$_{2}$\cite{EuFeAs1,EuFeAs2,EuFeAs3,EuFeAs4,EuFeAs5}. This assertion is further supported from density functional calculations on the superconducting CeO$_{0.5}$F$_{0.5}$BiS$_{2}$, which suggest that the rare-earth band is isolated away from the Fermi Surface\cite{Saxena}. In the CeFeAs$_{1-x}$P$_{x}$O system, for comparison, heavy fermion behavior at x $>$ 0.9 is explained by inter-layer hybridization between Fe-$3\mathit{d}$ and Ce-$4\mathit{f}$ electrons, and reduction of this coupling leads to the onset of ferromagnetism\cite{CeTMPnO4}. Recent Ce L$_{3}$-edge XAS experiment on CeO$_{1-x}$F$_{x}$BiS$_{2}$ also suggested reduced hybridization between the Ce-$4\mathit{f}$ orbital and Bi-$6\mathit{p}$ conduction band in the superconducting F-doping region\cite{XAS} due to structural displacement of S ions. The authors claimed that the reduced hybridization can be responsible for onset of both ferromagnetism in the Ce(O,F)-layer and superconductivity in BiS$_{2}$-layer. It should be noted, however, that it is also possible that the pairing symmetry is unconventional as in a spin-triplet pairing\cite{Triplet} enabling a mutually supportive coexistence between them. The pairing state of superconductivity in this system remains unclear and calls for further studies.

\section{Conclusion}

We have studied the nature of magnetism in the rare case of a ferromagnetic
superconductor in CeO$_{0.3}$F$_{0.7}$BiS$_2$. The neutron powder
diffraction shows broadening of the nuclear Bragg peaks suggesting an imperfect
crystalline structure as in other compounds in the new BiS$_2$-based
superconducting family\cite{JSLee,Annealing}. The magnetic structure
investigation showed that the magnetic moments are aligned ferromagnetically along the c-axis below T$_M$ $\approx$
7 K. Upon application of an external magnetic field, the ferromagnetic component develops in the $ab$-plane. From inelastic neutron
scattering measurements, it is shown that the Ising-like spin Hamiltonian with NN and
NNN interactions can best describe the observed spin-wave near E $\approx$
1.8 meV. An external magnetic field splits the spin-wave into two modes: one
excitation increases in energy while the other remains nearly at the same
energy. The anomalous phenomenon observed under field is discussed
in terms of CEF and asperomagnetism, both of which are related with an
Ising-like spin nature. In this new ferromagnetic superconductor, there
seems to be little interaction between the ferromagnetic and
superconducting layers.

\section{Acknowledgment}

The work at the University of Virginia has been supported by the National
Science Foundation, Grant number DMR - 1404994. \ A portion of this research at ORNL's High Flux Isotope Reactor and Spallation Neutron Source was sponsored by the Scientific User Facilities Division, Office of Basic Energy Sciences, U.S. Department of Energy.


\end{document}